\documentclass{aa}
\usepackage{graphicx}
\begin{document}
\title{Numerical simulation of the surface flow of a companion star in a
  close binary system}

\author{Kazutaka Oka\inst{1}, Takizo Nagae\inst{1}, Takuya Matsuda\inst{1},
  Hidekazu Fujiwara\inst{2}
  \and
  Henri M.J. Boffin\inst{3}
  }

\institute{Department of Earth and Planetary Sciences, Kobe University, 
  Kobe 657-8501, Japan\\
  \email{kazutaka@kobe-u.ac.jp, nagae@kobe-u.ac.jp, tmatsuda@kobe-u.ac.jp}
  \and
  IBM Japan Ltd., Yamato-shi, Kanagawa 242-8502, Japan
  \and
  Royal Observatory of Belgium, 3 av. Circulaire, 1180 Brussels, 
  Belgium  \\
  \email{Henri.Boffin@oma.be}
  }

   \date{Received 29 April 2002/ Accepted 12 July 2002}

\abstract{
  We simulate numerically the surface flow of a gas-supplying companion star 
  in a semi-detached binary system. Calculations are carried out for a region
  including only the mass-losing star, thus not the mass accreting star. The
  equation of state is that of an ideal gas characterized by a specific heat
  ratio $\gamma$, and the case with $\gamma=5/3$ is mainly studied. 
  
  A system of eddies appears on the surface of the companion star: an eddy
  in the low pressure region near the L1 point, one around the high
  pressure at the north pole, and one or two eddies around the low
  pressure at the opposite side of the L1 point.
  Gas elements starting near the pole region rotate clockwise around the north
  pole (here the binary system rotates counter-clockwise as seen from the north
  pole). Because of viscosity, the gas drifts to the equatorial region, switches
  to the counter-clockwise eddy near the L1 point and flows
  through the L1 point to finally form the L1 stream. 
  
  The flow field in the 
  L1 region and the structure of the L1 stream are also considered.
  
  }

\titlerunning{Numerical Simulation of the Surface Flow of a
  Companion Star}
\authorrunning{Oka et al.}
\maketitle
%

\section{Introduction}

A semi-detached binary is a system consisting of a primary star and a
Roche-lobe filling secondary star. In cataclysmic variables (CVs), for instance, the primary star
is a white dwarf, while the Roche-lobe filling star is a low-mass main-sequence star, 
from which gas is supplied to the primary star via an accretion
disc (see Warner 1995 for a review).

Accretion discs have been the main target of astrophysical research, while
gas supplying companion stars has not been investigated fully from both
theoretical and observational viewpoints. 
Knowledge of the properties of the companion star is required in order to
understand the evolutionary mechanism of the binary systems.
One reason that the companions have
not attracted much attention from astrophysicists is that they had little
observable constraints. Recently, however, Dhillon and
Watson (\cite{dhi00}) have demonstrated the technique of imaging the companion stars
in CVs using Roche tomography. 

The companion stars in CVs are similar to low-mass main-sequence stars in
their gross properties. However, companion stars exist in a different
environment from those of isolated stars. They might be influenced by 
effects like irradiation from the accretion disc around the primary star, its
rapid rotation, Roche-lobe shape, mass loss, and so on. These effects could in
turn affect the evolution of the binary systems. 

\subsection{Astrostrophic wind of Lubow and Shu}

Lubow and Shu (\cite{lub75}, \cite{lub76}) conducted pioneering work 
in the study of gas dynamics in semi-detached binaries. In their paper, they analyzed
the surface layer of the companion star, as well as the L1 region
(Lubow and Shu \cite{lub75}) and the L1 stream (Lubow and Shu \cite{lub76}).
In their analysis they solved the hydrodynamic equations by a matched 
expansion technique and reduced partial differential equations to ordinary 
differential equations.  
Their method was superior to numerical methods at that time 
considering the poor capability of computers.  
Since then, the progress in computer power has been considerable, so that it is worthwhile to 
investigate the same problem numerically. 
Our main target in the present paper is 
to investigate the surface layer, the L1 region and the L1 stream numerically.
The region around the accreting compact star has been investigated separately by our group 
(Makita, Miyawaki and Matsuda \cite{mak00}, Matsuda et al. \cite{mat00}, 
Fujiwara et al. \cite{fuj01}).

Lubow and Shu (\cite{lub75}) stated that ``the horizontal component of the 
fluid velocity is parallel to isobars in the astrostrophic approximation. 
Like the large-scale circulation patterns in the atmosphere and oceans of the 
Earth, the flow does not simply proceed from high pressures to low, 
but is directed around the contours of equal pressure.
In the northern hemi-lobe, the flow proceeds by keeping the region of high
(low) pressure to the right (left); in the southern hemi-lobe, 
by keeping high (low) pressure to the left (right).''
On a given equi-potential surface they expected ``the pressure to be 
largest at the poles (for the same reason that meridional circulation occurs
in rotating stars) and smallest at L1 (because of the mass-loss flow)''; 
therefore ``the excess circulation should be counter to the sense of the orbit 
rotation.''

They predicted that ``on each equipotential surface, the flow may be
astrostrophic, i.e., parallel to isobars, but a net transfer of  
matter within the entire surface layer to the equator is possible 
... if there is an outward flux of matter at the bottom of the surface layer.''
They also considered that ``another effect which reinforces the concept 
that there must be a slow drift of matter from the poles to the equator, and 
thence to L1, occurs if the envelope of the contact component is convective. 
In this case, the effective frictional drag exerted on the upper layers by 
the lower layers must produce the ``Ekman effect" (Ekman \cite{ekm22}) where
the induced secondary flow corresponds to a drift across isobars from high 
pressures to low.''

One of the aims of the present paper is to investigate the flow pattern 
of the gas in the surface layer of the companion numerically and to compare 
the results with the qualitative prediction of Lubow and Shu (\cite{lub75}). 
This kind of problems may be termed {\it stellar meteorology}.
The structure of the L1 region and the L1 stream are also considered.

\subsection{Our previous work}

Two-dimensional finite-volume hydrodynamic simulations of flows in a 
semi-detached binary were done by Sawada, Matsuda and Hachisu (\cite{saw86},
\cite{saw87}). Three-dimensional calculation was done by Sawada and Matsuda
(\cite{saw92}). They were concerned mainly by the accretion disc and did not
pay much attention to the surface flow on the secondary companion. 

Recently, Fujiwara et al. (\cite{fuj01}) performed three-dimensional
hydrodynamic simulations of a semi-detached binary system including 
the companion star. The equation of state is that of an ideal gas 
characterized by the specific heat ratio
$\gamma$ and the case with $\gamma=1.01$, meaning that the gas is almost
isothermal, was mainly studied. The reason for the choice of $\gamma=1.01$ is
to mimic a radiative cooling effect occurring in an accretion disc. Although
their main concern was the hydrodynamics of the accretion disc, they also
obtained the flow pattern on the surface of the companion.  Their results
showed that gas flows from the pole region to the equatorial region rather
directly, although small eddy patterns appeared on the surface of the
companion star. One, however, has to be careful to select the values of the
specific heat ratio. In the surface region of the companion, where the gas is
expanding outward, the temperature of the gas is reduced due to the adiabatic
expansion. Thus the case with $\gamma=1.01$, which results in the input of energy
into the expanding gas, may not be appropriate for the present purpose.

This paper is organized as follows. In ${\S}$ 2 we describe the numerical
method and physical assumptions. In ${\S}$ 3 we show the results of our
numerical simulations. In ${\S}$ 4 we discuss the dynamics of the obtained
eddy system. A summary is given in ${\S}$ 5.


\section{Assumptions and Numerical Method}

\subsection{Assumptions}
We consider the present study as a first step in the investigation of the surface flow
of the companion, and therefore do not take into account such complex effects
as viscosity (except numerical), magnetic fields, and irradiation from
the accretion disc. The equation of state considered here is that of an ideal
gas characterized by a specific heat ratio $\gamma$, and the cases with
$\gamma=5/3$ and 7/5 are mainly studied, although the cases of $\gamma=1.2$ and 1.01
are also considered to compare the present results with those of Fujiwara 
et al. (\cite{fuj01}).

In the present study, the binary system consists of a primary star with mass
$M_1$ and a companion star with mass $M_2$ rotating counter-clockwise when viewed
from the north pole. The mass ratio of the two stars, $q=M_2/M_1$, is assumed
to be unity in the present work. We normalize all physical
quantities as follows: the length is scaled by the separation $A$ between the
centres of the two stars. The time is scaled by $1/\Omega$, where $\Omega$ denotes
the orbital angular velocity, and therefore the orbital period
becomes $2\pi$. The density at the inner boundary is taken to be unity. 
The gravitational constant $G$ is eliminated using the above normalization. 

\subsection{Numerical method}
We use a Cartesian coordinate system. The origin of the coordinate is located
at the centre of the companion star. The $x-$axis coincides with the line
joining the centre of the two stars. We set the $x-y$ plane as the orbital plane,
thus the $z-$axis is perpendicular to the orbital plane and is oriented in the
same direction as the angular momentum vector of the orbital rotation. The L1
point is located at ($0.5, 0, 0$) and the primary star is located at 
($1,0,0$), which is out of the present computational region. 

The computational region is a rectangular box of size $-0.5<x<0.7$,
$-0.5<y<0.5$ and $0<z<0.5$. 
Note however that part of this box contains the mass-losing star, for 
which we assume that the inner part contains uniform density 
gas at rest (see Sect. 2.3).
With the assumption of symmetry of physical
quantities around the orbital plane, the calculations are performed only in
the upper half region above this plane. This region is divided in
$121{\times}101{\times}51$ grid points. In order to test the assumption of
symmetry about the orbital plane, we performed a test in which the whole space 
was taken into account in the computation. We confirmed that the assumption
of the symmetry was good enough.

In the present study, we use the same scheme as in Makita, Miyawaki, and
Matsuda (\cite{mak00}), Matsuda et al. (\cite{mat00}) and 
Fujiwara et al. (\cite{fuj01}): the simplified flux splitting (SFS) finite-volume 
method proposed by Jyounouchi et al. (\cite{jou93}) and Shima and
Jyounouchi (\cite{shi94}). We refer to Makita, Miyawaki, and  Matsuda (\cite{mak00})
for the numerical and the SFS schemes. With this MUSCL-type technique,
we can keep the spatial and temporal accuracy at second-order levels.

\subsection{Boundary and initial conditions}
We apply the same boundary conditions as in 
Fujiwara et al. (\cite{fuj01}) except for the inner boundary location of the 
companion star. The inner boundary
is assumed to be an equi-potential surface slightly smaller than the critical
Roche lobe: the mean radius of the inner boundary is chosen to be 0.45 in our
dimensionless unit. Note that this inner boundary does not necessarily mean
the real surface of the companion, but is chosen only for numerical purposes.
We have also made simulations where the inner boundary was chosen to be
0.40 and 0.50 and the results are very similar. In the rest of this paper, 
we will thus concentrate on the case with an inner boundary of 0.45.

What we call here the ``{\it surface}'' of the companion star might not correspond to
the photosphere, where optical depth reaches unity. Lubow and Shu 
(\cite{lub75}) pointed out that
``the depth over which the optical depth at visual wavelengths reaches unity
is not a {\it priori} related to the depth over which appreciable mass flow 
occurs.'' In the present study, we are interested in the surface flow of the
companion, and therefore we define the ``{\it surface}'' as the region where 
appreciable flow occurs. 

The inside of the companion star is filled with gas having zero velocity,
density $\rho_0=1$, and sound speed $c_o$, where $c_o$ is a free parameter. We
assume $c_0=0.05$ in the present work, although a case with $c_0=0.1$ is also
calculated and does not show much difference in the results. 
Such a value for the sound speed might, at first glance, look very 
large, if one would only take into account the effective temperature of the 
star and scale the corresponding speed with the orbital velocity. This is
however not correct because as we noted above, the inner boundary is not 
the "{\it surface}" of the star but may well lie more inside the star, where 
the gas is much hotter. Temperatures of several 10$^5$ K corresponding to 
sound speeds of the order of 50 km/s or more are therefore not unrealistic.

The assumption that the gas inside the companion star has zero
velocity does not mean that there is no gas outflow from the companion
star. The gas does flow out when the pressure above the inner boundary is
lower than that inside the companion. The outflow flux is calculated by solving 
Riemann problems at each step.

The outside of the outer boundary is assumed to be always filled with a gas
with velocity $0$, density $\rho_1$, and sound speed $c_1$, where $\rho_1$ and
$c_1$ are parameters. Use of these boundary conditions ensures a stable
calculation. Note that inflow or outflow of the gas through the outer boundary
is also possible, and the flux is calculated by solving Riemann problems between
the inside and outside of the computational region at each step. 

At the initial time $t=0$, the entire region, except the inside of the
companion, is occupied by a gas with velocity $0$, density $\rho_1=10^{-5}$, 
and sound speed $c_1=\sqrt{10}$.

To know how gas elements from the companion build the L1 stream is 
interesting, because this stream does bring the material from the companion to
the primary via the L1 point and an accretion disc. Thus we investigate the
structure of the L1 stream as well as the surface flow in the present work.

Simulations are run up to $t=62.8$ ($10$ orbital periods) except for the finer
grid case, in which the simulation is run up to $t=12.56$. We find that
the main flow patterns reach a steady state by these times.

\begin{figure*}
  \centering
  \includegraphics[width=10cm]{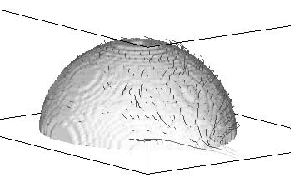}
  (a)
  \includegraphics[width=10cm]{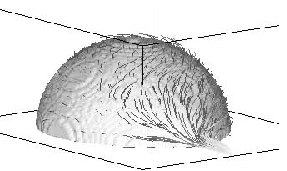}
  (b)
  \includegraphics[width=10cm]{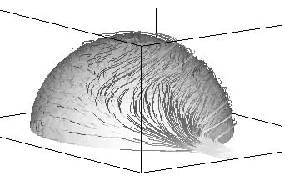}
  (c)
  \includegraphics[width=10cm]{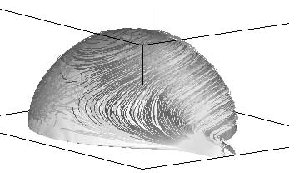}
  (d)
  \caption{Equi-density surface of the companion star and streamlines
    starting from the equi-density surface. Integration time steps to calculate
    the streamlines are the same for all streamlines, and so the shorter the lines
    the slower the speed of gas. Note that the equi-density surface is not
    completely opaque. The equi-density surface are defined by
    (a) $\log_{10}\rho=-1$; (b) $-1.5$; (c) $-2$; (d) $-2.5$.}
  \label{fig1}
\end{figure*}
\begin{figure}
  \centering
  \resizebox{\hsize}{!}{\includegraphics{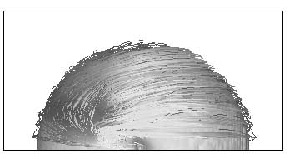}}
  \resizebox{\hsize}{!}{\includegraphics{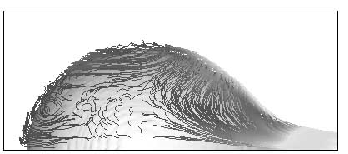}}
  \resizebox{\hsize}{!}{\includegraphics{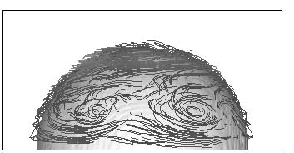}}
  \resizebox{\hsize}{!}{\includegraphics{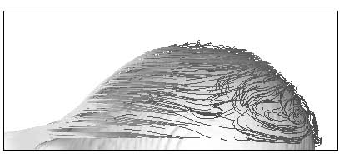}}
  \caption{Streamlines starting from the equi-density surface of
    $\log_{10}\rho =-2.5$ and equi-density surfaces of $\log_{10}\rho =-2$ 
    seen from various view angles.}
  \label{fig2}
\end{figure}
\begin{figure}
  \resizebox{\hsize}{!}{\includegraphics{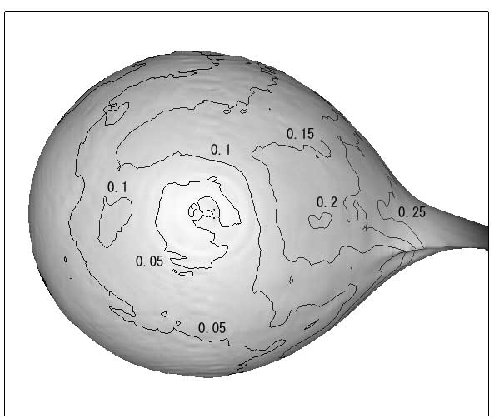}} 
  \caption{Iso-velocity lines plotted on the equi-density surface of 
    $\log_{10}\rho=-2.5$ seen from the north pole. The gray color
    shows the combined effect of the magnitude of the velocity and the 
    lighting effect.}
  \label{fig_vel}
\end{figure}

\section{Numerical results} 

\subsection{Astrostrophic wind}
    
Figures \ref{fig1}(a)-(d) show the equi-density surface of the companion star 
and the streamlines starting from the equi-density surface; Fig. \ref{fig1}(a)
depicts the equi-density surface of $\log_{10}\rho=-1$, (b) $-1.5$, 
(c) $-2$, and (d) $-2.5$. 
In order to draw the streamlines, we use the same integration time step. 
Therefore, the higher the density of gas, the shorter the streamlines 
as the velocity is lower. We observe that there are ascending 
motions of gas
in the higher density region (deeper inside the star) and a remarkable
circulating flow in the lower density region (surface region). The
streamlines of lighter color have higher velocity than those of darker ones. 
As will be shown later, there is a high pressure around the north pole, and 
so we may call this circulatory flow the north pole circulatory flow or 
simply an H-eddy. 

In order to see the surface flow more closely, Fig. \ref{fig2} shows the 
equi-density surface ($\log_{10}\rho=-2$) of the companion star and the 
streamlines starting from the equi-density surface of $\log_{10}\rho=-2.5$ 
seen from various view angles. We confirm the presence of the H-eddy again. 
Gas in the H-eddy seems to drift gradually downward and its velocity increases 
with decreasing latitude. This phenomenon will be discussed later.

Figure \ref{fig_vel} shows the iso-velocity lines plotted on the equi-density 
surface of $\log_{10}\rho=-2.5$ seen from the north pole. The numbers in the
figure show the magnitude of the velocity (note that the velocity is 
normalized by $A\Omega$, where $A$ is the separation and $\Omega$ is the 
rotational frequency of the binary.)
 
In the L1 region, which is close to the L1 point, there is a strong low
pressure due to the mass outflow through the L1 point. Because of the presence
of the L1 low pressure, there is a circulatory flow, around the L1 point, 
rotating counter-clockwise, although this eddy does not form a complete circle
for geometrical reasons. We call this flow the L1-eddy, 
although, strictly speaking, this is not a complete eddy.

The other remarkable feature is the presence of two eddies rotating
counter-clockwise at the opposite side of the L1 point. These two eddies merge
to form one larger eddy at different density levels. Since this region is
close to the L2 point, we may call this eddy the L2 eddy. Since these H, L1
and L2 eddies are explained in terms of the astrostrophic wind
(Lubow and Shu \cite{lub75}), the pressure distribution on the surface of the 
companion is important.

\begin{figure}
  \resizebox{\hsize}{!}{\includegraphics{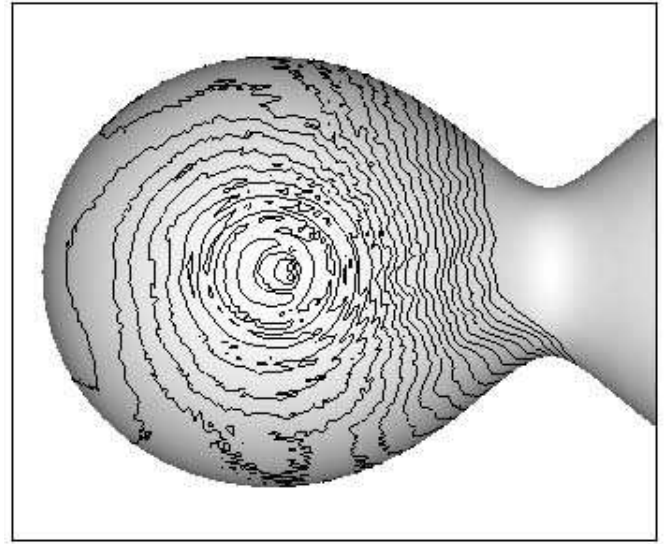}} 
  \caption{Isobaric lines plotted on the equi-potential surface seen from
    the north pole.}
  \label{fig3}
\end{figure}

\begin{figure}
  \resizebox{\hsize}{!}{\includegraphics{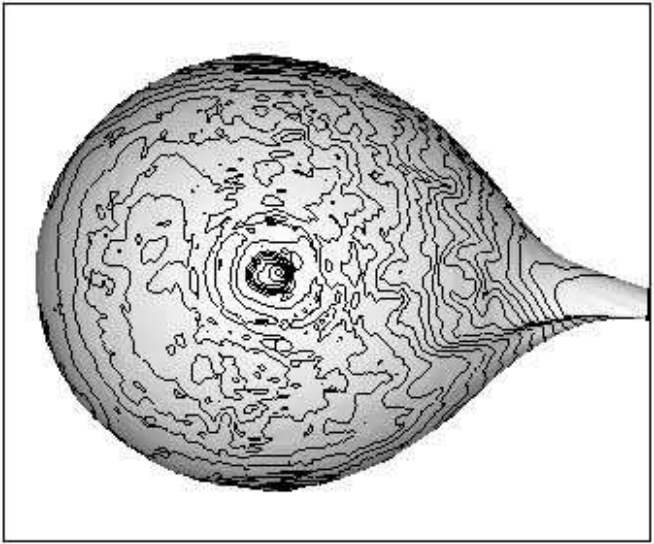}} 
  \caption{Isobaric lines plotted on the equi-density surface of 
    $\log_{10}\rho=-2.5$ seen from the north pole.}
  \label{fig4}
\end{figure}

\subsection{High / low pressures on the companion surface}

If there is no flow on the companion, the isobaric surface should coincide
with the equi-potential surface as a result of the hydrostatic balance.  In our
case, the isobaric surface does not coincide with the equi-potential
surface in an exact sense because of the presence of the surface flow. 
Fig. \ref{fig3} shows the
isobaric lines plotted on the equi-potential surface of the companion. This
equi-potential surface nearly corresponds to the equi-density surface with
$\log_{10}\rho=-2.5$ except near the L1 region, where high speed gas motion 
is present. Note that the isobaric lines on the companion surface have
concentric  patterns around the north pole.

As was described earlier, the flow pattern on the surface of the rotating (in
the inertial frame) companion can be explained in terms of the astrostrophic
wind, which is a counterpart of the geostrophic wind in the atmosphere of the
Earth. If viscosity can be neglected, the direction of the flow is parallel to
the isobaric lines of low/high pressure. In the present case of a
counter-clockwise rotating binary system (as viewed from the north pole), the 
flow proceeds by keeping the high (low) pressure to the right (left) in the 
northern hemisphere.
Therefore, the H-eddy circulates clockwise, while the L1 eddy and L2 eddy 
circulate in the counter-clockwise direction. 

If the pressure is only a function of the density of gas, the gas
is called barotropic. Otherwise, the gas is called baroclinic. 
In the case of $\gamma=1.01$, the gas is nearly iso-thermal, hence almost barotropic.
In the other cases, the gas is baroclinic. In a barotropic gas, an isobaric surface 
must coincide with an equi-density surface. Fig. \ref{fig4} shows the 
baroclinicity of the gas with $\gamma=5/3$, whose property is responsible 
for the generation of vorticity. This will be discussed later.

\subsection{Ekman boundary layer}

In the Earth's atmosphere, an Ekman boundary layer exists close to the 
surface of the
Earth, where (turbulent) viscosity cannot be neglected and where the direction of the flow
deviates from that of a pure geostrophic wind. In low pressure regions, gas
drifts inwards and eventually ascends near the centre of the low pressure regions
(Pedlosky \cite{ped87}).

\begin{figure}
  (a)\\
  \resizebox{\hsize}{!}{\includegraphics{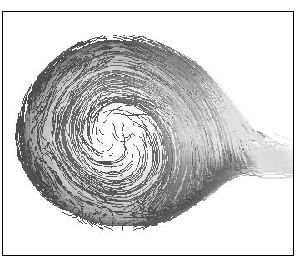}}
  (b)\\
  \resizebox{\hsize}{!}{\includegraphics{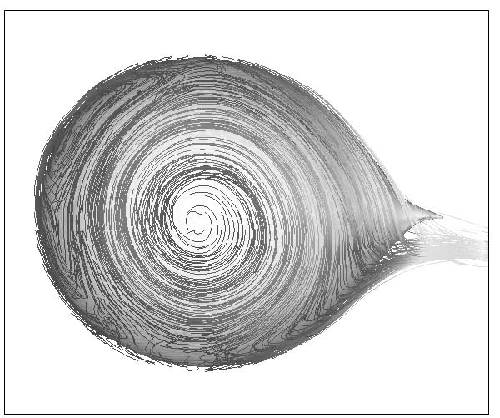}} 
  \caption{(a) Calculation with mesh size $121{\times}101{\times}51$. 
    (b) Calculation with mesh size $241{\times}201{\times}101$
    at $t=12.56$.}
  \label{fig5}
\end{figure}

In our simulations, there is no (physical) viscosity assumed. Nevertheless,
numerical viscosity arising from a truncation error acts as a kind of a
turbulent viscosity, leading to gas drifting.
This is the
reason why the circulating flow about the north pole gradually migrates
towards the equatorial region.  In order to test the above consideration, we
perform a calculation with the mesh size halved. In this case the magnitude
of the (numerical) viscosity is expected to be $1/8$ of the coarser
grid case. We find a similar flow pattern as before, although the circulating
motion is tighter than before (see Figs. \ref{fig5}). The main characteristics of the flow pattern 
are not changed, though. 

In the surface region of low-mass main-sequence stars, there is a convection
zone, which provides the necessary turbulent viscosity. We may therefore expect
that the present flow pattern is qualitatively correct. 

It is however not trivial from our simulations to derive the exact value of
our (numerical) viscosity. We can only make a very tentative guess that the 
typical Reynolds number in our 
simulations is probably of the order of 10 to 100. 
In a future work, we plan to solve this problem using the Navier-Stokes equation
instead of the Euler ones, and using the molecular viscosity as a rough 
estimate of the turbulent viscosity which should be at play here.

In our case the inner boundary does not play the role of the surface of the
Earth. The Ekman layer is present at the top of the atmosphere, a
situation similar to a water vortex. 
Gas circulating around the high pressure region at the north pole migrates
gradually towards the equator, switches to the counter-clockwise motion
close to the L1 low pressure, and finally escapes from the companion towards the
primary star. This flow pattern is important to understand the structure of the
L1 stream to be discussed later.

\subsection{Effect of $\gamma$}

Until now we have shown the results for $\gamma=5/3$, corresponding to ionized 
hydrogen gas. In order to study the case of molecular hydrogen, we calculate 
the case $\gamma=7/5$. The results are essentially the same as 
for $\gamma=5/3$, so that we do not discuss them any further.

However, the results for $\gamma=1.01$ are very different from
the cases of $\gamma=5/3$ and 7/5, and are essentially the same as that of Fujiwara
et al. (\cite{fuj01}): there is no H eddy but the flow originating at the pole
region migrates rather directly towards the equatorial region. More precisely,
such a flow is rather weak, because the pressure on an equi-potential
surface is rather uniform except in the region close to the L1 point. The L1 flow
mainly originates from the region close to the L1 point. 
The case of $\gamma=1.2$ is intermediate to the above two extreme cases. 
We found the L2 eddy in this case.

We may explain the difference of the flow patterns between the cases of 
$\gamma=5/3, 7/5$ and that of $\gamma=1.01$ in terms of the baroclinicity of
the gas. Vorticity is generated by the term containing 
$\nabla \rho \times \nabla p$, which is identically zero for a barotropic 
gas (Pedlosky \cite{ped87}). Therefore, local vorticity of the gas in the 
case of $\gamma=1.01$ is almost zero everywhere. Moreover, the
  Taylor-Proudman theorem, which states that the horizontal velocity
  components $u$ and $v$ are independent of $z$, holds not only for 
an incompressible flow but also for a barotropic gas 
(Busse \cite{bus70}; Pedlosky \cite{ped87}). Since $u$ and $v$ are almost zero
on the inner boundary, they must also be small in the upper atmosphere.
 This problem, however, needs
further investigation. We note that
the cases of $\gamma=5/3$ and 7/5 are more appropriate to compare with (future) observations.

\subsection{L1 region and L1 stream}

Lubow and Shu (\cite{lub75}) paid lots of effort to solve the flow in the L1 
region by their matched expansion technique. In this subsection we summarize
our numerical results. 

In order to make the influence of the outer boundary to the L1 stream small, 
we slightly expand the computational region of the $x$-component from $0.7$ 
to $0.9$ in this subsection.
The primary star is 
still out of the computational
region. Thus the L1 stream reaches the boundary at $x=0.9$ and 
escapes from the calculation region. In the present calculation no accretion
disc is formed, and so there is no collision between the L1 stream and the
accretion disc around the primary star.

\begin{figure}
  \resizebox{\hsize}{!}{\includegraphics{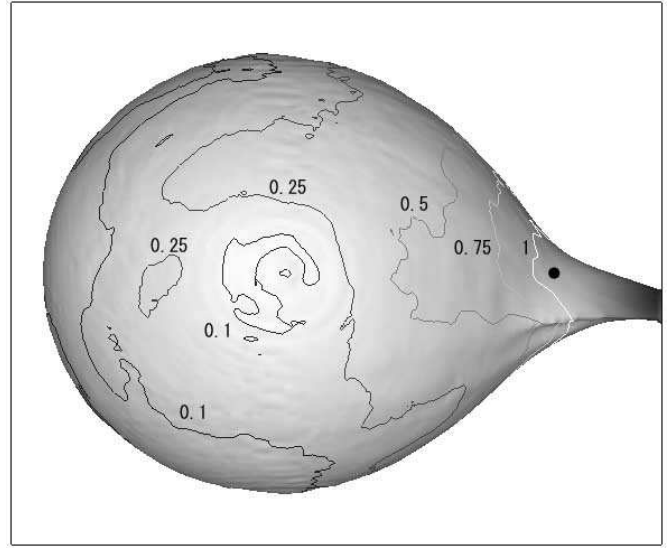}} 
  \caption{Iso-Mach number lines on an equi-density surface with
    $\log_{10}\rho=-2.5$ seen from the north pole. The sonic line is 
    close to the L1 point. The gray color shows the combined effect of the 
    magnitude of the velocity and the lighting effect. The heavy black 
    dot shows the position of the inner L1 point.}
  \label{fig_mach}
\end{figure}

Lubow and Shu (\cite{lub75}) predicted that the sonic point, where subsonic gas
is accelerated to supersonic velocity, must occur close to the L1 point.
Figure \ref{fig_mach} shows the iso-Mach number lines on an equi-density 
surface and the numbers on the figure show the Mach number. 
It can be seen that the sonic line on the surface is certainly close to the L1 
point.

  \begin{figure}
   (a)\\
   \resizebox{\hsize}{!}{\includegraphics{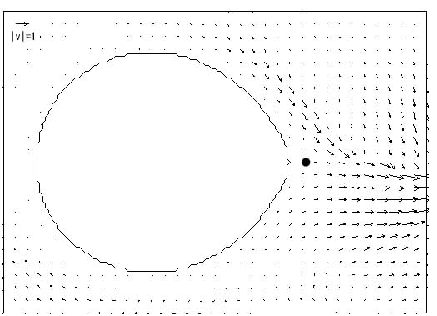}}\\
   (b)\\
   \resizebox{\hsize}{!}{\includegraphics{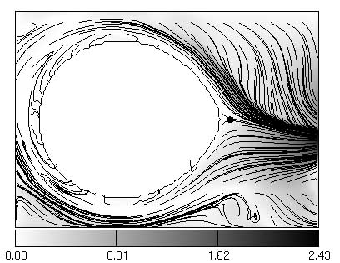}}\\
   (c)\\
   \resizebox{\hsize}{!}{\includegraphics{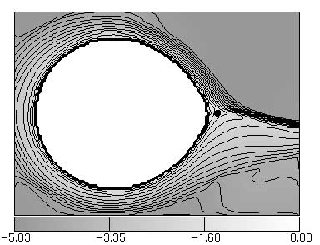}} \\
%
%
  \caption{(a) Velocity vector in the orbital plane.  Note that arrows
are drawn only every four cells in each x and y direction.
    (b) Streamlines in the orbital plane with velocity magnitude
    contours. (c) Density contours with equi-density lines. The heavy black 
    dot shows the position of the inner L1 point.}
  \label{fig7}
%
  \end{figure}

Figs. \ref{fig7}(a)-(c) show the velocity vectors, the streamlines, and 
the equi-density lines of the gas in the orbital plane, respectively.
The flow pattern in the orbital plane near the
L1 point shows some difference with the one depicted in Fig. 3 in Lubow and Shu (\cite{lub75}). They
predicted that the streamlines of the lower portion of the L1 point turn
back to the companion, while in our case all gas flow towards the primary
compact component. There are two differences between thier assumptions
and ours: 1) they assumed iso-thermal gas; 2) they start their calculation
at the L1 point, where the gas is assumed to be at rest. 
The assumption 2) may be the main reason. 
We solve the whole region including both the L1 region and the surface layer 
of the companion. The gas at the L1 point is already supersonic as 
was described above.

From Fig. \ref{fig7}(b) we can measure the deflection angle $\theta_s$ of 
the L1 stream, which is the angle between the central flow line in the L1 
stream and the $x-$axis. It is about -10 degrees. 
This angle was calculated to be about 
-20 degrees for the case of $q=1$ by Lubow and Shu (\cite{lub75}). In
Fujiwara et al.'s result (\cite{fuj01}) this is about -15 degrees.
This difference is also due to the difference in boundary conditions.
Lubow and Shu assumed that the fluid element was at rest at the L1 point.
The effect of the Coriolis force is measured by the Rossby number defined by  
$U/A\Omega$, where $U$ is a typical velocity of gas. The lower 
the initial velocity of the gas, the lower the Rossby 
number, the larger the Coriolis' effect and, hence, the deflection angle. 

In Fig. \ref{fig7}(c) we found that the atmosphere of the secondary is 
thicker in the leading hemisphere (negative $y$) than the trailing hemisphere.
This can be easily explained by the direction of the Coriolis force, 
which acts to the right side of the flow direction in our system.

\begin{figure}
  (a)\\
  \resizebox{\hsize}{!}{\includegraphics{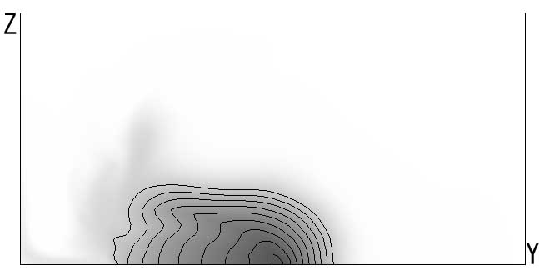}} 
  (b)\\
  \resizebox{\hsize}{!}{\includegraphics{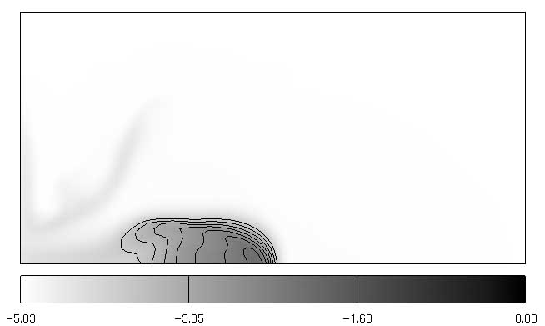}}
  \caption{Density contours of the L1 stream at two cross sections: 
    (a) $x=0.5$; (b) $x=0.65$. The horizontal axis is the $y$-axis
    ranging from -0.5 to 0.5. The vertical axis depicts the $z$-axis ranging from 
    0 to 0.5. Note the asymmetry of the density contour in the central 
    part of the L1 stream.}
  \label{fig8}
\end{figure}

\begin{figure}
  \resizebox{\hsize}{!}{\includegraphics{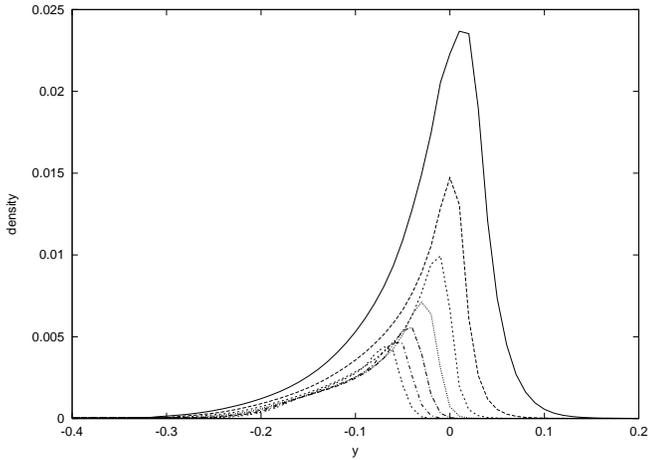}} 
  \caption{Density profile along the $y$-axis on the orbital plane 
    $z=0$ at various $x$
    locations, ranging from $0.5$ to $0.8$.}
  \label{fig}   
\end{figure}

Because of this asymmetry of the atmosphere, the structure of the L1 stream
is also asymmetric about the central flow line.
In Figs. \ref{fig8} we show the cross section of the L1 stream, which is bow
shaped convex to the positive $y$ direction. We obtained a similar pattern in
Fujiwara et al. (\cite{fuj01}), and this was explained by the 
interaction of the L1 stream and the circulating circum-disc flow on the 
orbital plane about the primary.  
In the present work we realize that this explanation is not correct. 
The bow shape of the L1 stream should be explained by the structure of the surface flow on 
the companion. 

\begin{figure}
  \resizebox{\hsize}{!}{\includegraphics{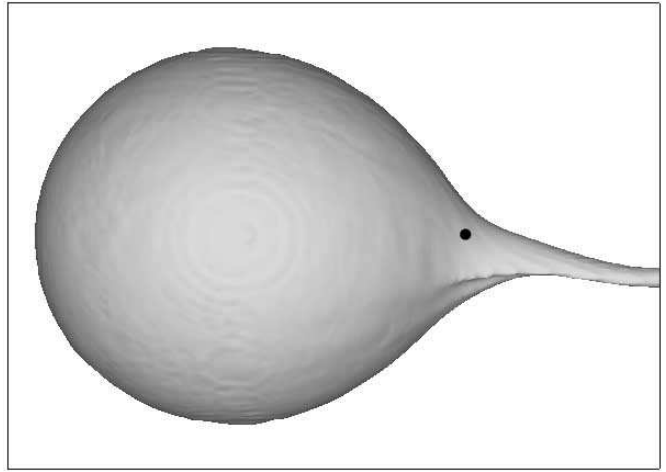}} 
  \caption{Equi-density surface of $\log_{10}{\rho}=-2.5$ of the companion 
    star and the L1 stream seen from the north pole. The heavy black 
    dot shows the position of the inner L1 point.}
  \label{fig9}
\end{figure}

Lubow and Shu (\cite{lub75}) predicted a symmetric Gaussian density 
distribution at the centre of the stream as
\begin{equation}
\rho(r, \theta){\approx}\frac{C}{r}\exp[-Br^2(\theta-\theta_s)^2],
\label{lub}
\end{equation}
where $r$ is the distance from the L1 point in the scaled 
coordinate, $\theta_s$ is the angle of the stream centre from the line 
joining the centre of the two stars, and $C$, $B$ are the coefficients 
depending on the mass ratio. 
Fig. \ref{fig} shows that our density profile nearly has a Gaussian density 
distribution. The density at the centre of the stream obtained by them 
decreases as $r^{-1}$, while our result decreases as $r^{-0.8}$. 
Fig. \ref{fig9} shows the equi-density surface of $\log{\rho}=-2.5$ of the 
companion star and the L1 stream.

\begin{figure}
  \resizebox{\hsize}{!}{\includegraphics{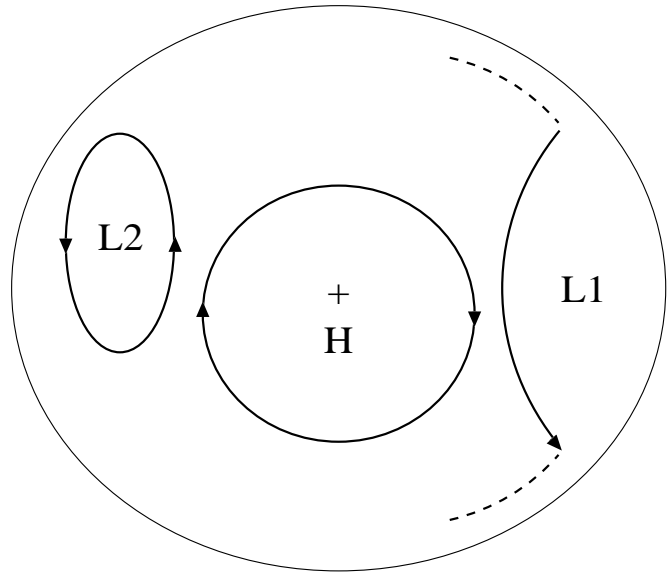}} 
  \caption{Schematic diagram of the system of eddies: L1, H, and L2 show the 
    low pressure near the L1 point, the high pressure around the north pole, 
    and the low pressure at the opposite side of L1, respectively.
    An eddy rotates clockwise for H, and counterclockwise for L1 and L2. }
  \label{fig10}
\end{figure}

\section{Discussion}

In our simulations, we obtained a system of eddies: the H eddy rotating clockwise, and 
counterclockwise rotating  L1 and L2 eddies. We now discuss what we believe is the reason 
that this eddy system
is formed (see Fig. \ref{fig10}). 

If we consider a time sequence, the L1 low pressure is formed first,
because of the mass loss through the L1 point. Since the gas near the equator does not
feel much resistance in flowing towards the L1 point, the equatorial region 
becomes a low pressure zone, and a high pressure zone builds up at the north pole: 
the H eddy is thus formed.

The above explanation does not explain the appearance of the third eddy: the
L2 eddy. This can be explained as follows. Consider a force exerted by an eddy
on another eddy.  The L1 eddy tries to convect the H eddy to negative
$y$-direction, while the H eddy tries to convect the L1 eddy to
negative $y$-direction as well. However, the L1 eddy must be fixed to the L1
region, so that the H eddy must drift to negative $y$-direction. The system is
therefore not steady. In order to stabilize the eddy system, we need the third eddy,
which tries to convect the H eddy to positive $y$-direction. The L2 eddy thus
appears and evolves until the three eddy system is stabilized. The exact location
and the strength of these eddies can be obtained only through numerical 
simulations.

It is interesting to note that
Davey and Smith (\cite{dav92}) have found that the trailing 
hemisphere has stronger line absorption than the leading hemisphere in a 
dwarf nova (see also Dhillon and Watson \cite{dhi00}). 
This asymmetry can be naturally explained in terms of the H-eddy current,
which convects heat due to irradiation to the leading hemisphere.

If this kind of astrostrophic wind conveys heat effectively from an
irradiated hemisphere to the other, the unirradiated side of the 
companion becomes hotter than previously thought.
Such an example may be observed in the 2000 outburst of the recurrent
nova CI Aql because, in nova outbursts, a companion star should be 
strongly irradiated by a hot white dwarf. 
The inclination angle of the CI Aql orbit is about 80 degrees and 
the eclipse is almost total in the sense that the companion completely
occults the bright accretion disc around the white dwarf  
(e.g., Hachisu \& Kato 2001).  In eclipse minima,
therefore, we see only the unirradiated side of the companion. 
Hachisu, Kato and Schaefer (2002) showed that the eclipse 
minima of CI Aql in outburst is about 0.6 mag brighter than in quiescence
and could reasonably reproduce the orbital light curve (folded 
by the orbital period) when the temperature of the unirradiated 
hemisphere is about 1000 K hotter than that in quiescence.

In the present study, we do not take into account such complex effects as
(real) viscosity, magnetic fields, and irradiation from the accretion
disc. Activities of magnetic fields on the companion surface would lead to
star-spots, and the irradiation from the accretion disc would lead to an increase
of the gas temperature of the companion. These effects might more or less change
the surface flow patterns. The present study is a first step to examine the
surface flow structure. Thus to study these phenomena is beyond the scope of the
present paper, and is left for future work.


\section{Summary}

We performed numerical simulations of the surface flow on the companion star
in a semi-detached binary system. The case with $\gamma=5/3$ was mainly 
studied, although the cases of $\gamma=7/5, 1.2$ and 1.01 were also investigated. 
The results are summarized as follows:
\begin{enumerate}
\item We obtain an eddy configuration composed of an H-eddy, an L1-eddy,
  and an L2-eddy associated with high/low pressures.  
  If there is no viscosity, the flow lines must be closed.
  The above eddies are explained in terms of the astrostrophic wind.

\item Some kind of viscosity (e.g., turbulent viscosity
  originating from a surface convection zone) would lead to a downward drifting
  motion. In our simulation the numerical viscosity acts as turbulent 
  viscosity.

\item The sonic line on an equi-density surface occurs close to the L1 point.

\item The deflection angle of the L1 stream is about -10 to -15 degrees.
  This small angle is due to the larger fluid velocity at the L1 point.

\item The density distribution in the L1 stream is nearly symmetric about the
  central flow line. The cross section of the L1 stream is bow shaped convex 
  to the positive $y$ direction. This is an intrinsic nature of the stream 
  and is not caused by the interaction with the accretion disc gas.
\end{enumerate}


\begin{acknowledgements}

The authors would like to thank Prof. Takahiro Iwayama for his valuable
discussions on meteorological aspects of the present study. 
The referee, Dr. S. Lubow, is gratefully acknowledged for valuable comments. 
T.M. was supported
by the grant in aid for scientific research of the Japan Society of Promotion of
Science (13640241). 
H.B. would like to thank the Kanbara-Tosao foundation for its financial
support. Calculations were carried out on SGI Origin 3800 at the 
Information Processing Center of Kobe University.

\end{acknowledgements}


\end{document}